# Non-perturbative Hydrodynamic Model for Multiple Harmonics Generation in Metallic Nanostructures

Pavel Ginzburg[1,2,=,*], Alexey V. Krasavin[1,=], Gregory A. Wurtz[1], and Anatoly V. Zayats[1]

[1]Department of Physics, King's College London, Strand, London WC2R 2LS, United Kingdom
[2]ITMO University, St. Petersburg 197101, Russia

**Abstract:** Optical response of free-electron gas leads to inherent nonlinear optical behaviour of nanostructured plasmonic materials enabled via both strong local field enhancements and inherent complex electron dynamics. We present a comprehensive treatment of microscopic polarization of conduction electrons in the time-domain using full hydrodynamic description which allows self-consistent modelling of linear and nonlinear response and multiple harmonics generation. The effects of convective acceleration, magnetic contribution of the Lorenz force, quantum electron pressure, and nanostructure's boundaries have been taken into account leading to simultaneous appearance of second and third harmonics. The developed method provides an ultimate approach to investigate nonlinear responses of arbitrarily-shaped complex nano-scale plasmonic structures and enables addressing their self-consistent nonlinear dynamics.

= Equal contribution
* Corresponding author: pavel.ginzburg@kcl.ac.uk

# 1. Introduction

Nonlinear optical interactions give rise to a variety of phenomena extensively used in numerous applications in lasers, classical and quantum optical information processing, bio-imaging and sensing. Consequently, tailoring, enhancing and controlling nonlinear processes which are inherently weak and demanding high light intensities, is the task of a prime significance. One of the promising approaches is to employ nanostructured materials to tailor both nonlinear susceptibilities and local electromagnetic fields.

Plasmonic nanostructures deliver both a true nanoscale light confinement and a strong nonlinear response, as both rely on peculiarities of the reaction of conduction electrons to electromagnetic field of incident light [1]. Both second and third harmonics generation, optical Kerr nonlinearity and possibility to achieve solitonic waves have been demonstrated in metal films and metallic nanostructures [2,3,4,5,6,7].

Typical theoretical approaches for description of nonlinear response of plasmonic materials rely on perturbative treatment of a nonlinear material polarization, either in the hydrodynamic model or phenomenologically, and enable separation of higher harmonics via quasi-Fourier transformation [8]. In contrast, the full time-domain analysis comprehensively addresses nonlinear dynamics without any additional assumptions on the interaction nature. Moreover, time-domain studies give opportunity to explore simultaneously both bulk and surface contributions to nonlinear generation processes, as well as efficiency of side bands generation, as will be shown hereafter.

Here, we develop a comprehensive and non-perturbative numerical model for investigation of nonlinear interactions of light with plasmonic nanostructures. The full time-domain analysis comprehensively addresses nonlinear dynamics of free electrons without any additional assumptions on the interaction nature and gives the opportunity to explore interplay between various nonlinear processes, and efficiency of side bands generation. Optical properties of metal particles are described with the help of a full hydrodynamic model [9], which enables to account for both linear and nonlinear dynamics of the conduction electrons under visible and infrared light illumination (away from the spectral range of the interband transitions). Hydrodynamic equations straightforwardly reproduce the Drude model in the linear regime of interaction with electromagnetic field. Additional effects, accounting for the response of core electrons [10] and employing quantum approaches for electron exchange correlations [11] may also be included. However, at higher intensities convective acceleration, magnetic contribution of the Lorenz force, and quantum electron pressure terms, present in the hydrodynamic model, introduce strong nonlinear contributions. Moreover, even at moderate intensities of illumination, resonant effects start to be especially important in small metal nanostructures, as they maximize nonlinear surface response [7]. All these effects may simultaneously be taken into account by coupling nonlinear hydrodynamic equations for electron plasma with the Maxwell's equations for electromagnetic fields [12]. As an example, nonlinear scattering on a gold nanocylinder was analyzed. Appearance of strong second-harmonic (classically forbidden, but enabled by boundary effects [13]) and third-harmonic signals were observed. The results are compared to phenomenological model.

## 2. Self-consistent formulation of electromagnetic-hydrodynamic problem

The interaction of electromagnetic fields with objects made from arbitrary (nonmagnetic) materials is described in terms of the induced polarization $\vec{P}$ via

$$\nabla \times \nabla \times \vec{E}(\vec{r},t) + \frac{1}{c^2}\partial_{tt}\vec{E}(\vec{r},t) + \mu_0 \partial_{tt}\vec{P}(\vec{r},t) = 0, \qquad (1)$$

where $\vec{E}(\vec{r},t)$ is the incident electrical field and $\mu_0$ is the vacuum permeability. In general, the coordinate-dependent polarization term contains all the information on both linear and nonlinear contributions, also including chromatic dispersion. In the framework of the hydrodynamic model, $\vec{P}$ is introduced via polarization currents, which are defined with the help of natural hydrodynamic variables – the macroscopic position-dependent electron density $n(\vec{r},t)$ and velocity $\vec{v}(\vec{r},t)$. The basic set of hydrodynamic equations is given by [e.g. 9]:

$$m_e n\left(\partial_t \vec{v} + \vec{v}\cdot\nabla\vec{v}\right) + \gamma m_e n\vec{v} = -en\left(\vec{E} + \vec{v}\times\vec{H}\right) - \vec{\nabla}p$$
$$\partial_t n + \nabla\cdot(n\vec{v}) = 0 \qquad (2)$$

where $m_e$ and $e$ are the electron mass and charge, respectively, $\gamma$ is the effective scattering rate, representing optical losses in a phenomenological way, and $p = \left(3\pi^2\right)^{2/3}\dfrac{\hbar^2}{5m_e}n^{5/3}$ is the quantum pressure which can be evaluated in the Thomas-Fermi theory of an ideal fermionic gas. The $\vec{v}\cdot\nabla\vec{v}$ term is the convective acceleration (in analogy to fluid dynamics) and is one of the key contributors to the nonlinear generation process. The electromagnetic (Eq. 1) and hydrodynamic (Eq. 2) sets of equations are coupled via the microscopic polarisation term:

$$\partial_t \vec{P} = -en\vec{v}. \qquad (3)$$

The set of Eqs. 1, 2 and 3 provides a self-consistent formulation of nonlinear optical processes due to free conduction electrons of in plasmonic structures. The effects of surface nonlinearities and nonlocality are taken into account in the above set of equations. The proposed approach is a new nonperturbative description of free-electron nonlinearities allowing taking into account a comprehensive hydrodynamic processes in the electron plasma. It is worth noting that, to the best of our knowledge, all previous studies addressed only linear single-frequency regimes while nonlinear contributions were taken into account perturbatively via quasi-Fourier transform techniques.

## 3. Results

Details of the numerical model

While the proposed method is universal and enables to address any geometry, here, as a particular example, the problem of nonlinear scattering of TM plane wave by infinitely long cylinder of nanoscale diameter is considered (Fig. 1(a)). This will allow comparison of our results to other known approaches developed for metal nanopartcticles.

The set of differential equations Eqs. 1-3 was numerically solved by employing the finite element method. A Gaussian pulse $\vec{E}_1(y,t) = (0, E_y) \cdot \exp[-y^2/(2w^2)] \cdot \exp[-(t-t_0)^2/(2\tau^2)] \cdot \cos[\omega_1 t]$ at fundamental harmonic frequency $\omega_1 = 1.257 \cdot 10^{15}$ rad/s (corresponding to the wavelength of $\lambda_1 = 1500$ nm) with a temporal width of $\tau = 20$ fs (and a spatial width of $w = \lambda/2$ was incident on an infinitely long cylinder of radius $r = 100$ nm. The pulse is incident along positive x-direction and polarisation was in y-direction (Fig. 1(b)). The time offset $t_0 = -3\tau$ and the simulation time span $T = 7\tau$ was chosen so that the scattered light pulse containing higher harmonics completely pass any point in 6 μm×6 μm simulation domain. It was additionally checked that any further increase of the simulation time-span does not affect the results. Tabulated physical constants were used for implementations of Eqs. 1-3, and the equilibrium free carrier concentration of gold was taken to be $5.98 \cdot 10^{28}$ [1/m$^3$] and $\gamma = 1.075 \cdot 10^{14}$[1/sec], and $\omega_p = 13.8 \cdot 10^{15}$[1/sec].

In the linear regime, Eqs. 2 and 3 provide general Drude-like response of the electron gas. This was numerically tested by comparing the low-intensity linear scattering in the full hydrodynamic model and the linear scattering simulation with the conventional Drude model (Fig. 3). Nonlinear terms in the full hydrodynamic model become significant under the high intensity excitation.

Nonlinear Spectrum

The time dependence of the scattered signal after the subtraction of the excitation pulse was probed in the near-field and the scattered optical signal, passing the boundaries of the simulation domain (far-field region) was Fourier-transformed into the frequency domain at each point of the boundary. Both near-field and far-field signals have the same spectra (Fig. 2(a)). The far-field spectral dependencies were integrated over the peaks at the second and the third harmonic frequencies, and, thus, the relative intensities of the nonlinear signal were found. Finally, the latter were integrated over the simulation domain boundaries and the total fluxes of the generated harmonics were obtained. As a crucial proof of the validity of the model, it was found that they follow the well-defined powers of the pump (quadratic and cubic for second and third harmonics, respectively), as expected from the theory of nonlinear optical processes.

With increase of the excitation intensity, the nonlinear harmonics grow as expected (the highest simulated intensity $I_0 = 9 \cdot 10^{18}$ W/m$^2$ corresponds to the intensity the particle can still withstand [14] with the correction for the incident light wavelength; please note that the temperature effects were not included in the model). The spectral widths of the nonlinear signals are two and three times broader than of the pump pulse. This is a direct result of the nonlinear processes since the nonlinear polarizations are narrower in time compared to the linear one, and thus has a wider frequency spectrum. The intensity of nonlinear scattering grows with certain proportionality to the intensity of the excitation (Fig. 2 b,c). In the chosen geometry, we can identify the main contribution to this term from the convective acceleration and the Lorentz force. The relative contributions of the various terms may be different depending on a particular geometry.

Second harmonic

Frequency domain analysis of hydrodynamic equations allows deriving surface polarizabilities [7] of a particle under undepleted pump assumptions. Effective nonlinear surface polarizability can also be introduced phenomenologically and related to the experimental data. The above approaches were adopted for studies of nonlinear (second harmonic) scattering from nanoparticles. Radiation patterns, calculated in the quasistatic limit [12,15,16], as well as complete solutions, relying on extended Mie theory [17, 18], were intensively studied. Advanced numerical modelling (yet in the frequency domain) enables to address more complex particle geometries [19]. One of the central points of the above studies was to account for nonlocal and retardation effects, responsible for the radiation pattern formation.

In order to compare our approach to the previous models, we have studied the second-harmonic field distribution calculated with the help of our full hydrodynamic approach (Fig. 3(a)). The field pattern was calculated by spectral frequency filtering and represents the intensity of the signal at each simulation domain point integrated over the spectral peak at the second harmonic frequency. It possesses a quadrupole characteristic with a dipolar contribution. Since the particle is made of centro-symmetric material and is also symmetric by itself, the dipolar radiation is forbidden by the selection rules. However, retardation effects of the electromagnetic wave propagating over the particle dimensions result in appearance of weak dipolar second-harmonic generation. The quadrupole, being significant contributor for the far-field pattern, originates from nonlocal interactions of electron plasma with particles boundary. While, this type of effect in the first approximation could be described with local polarization term of the form $P_{surf,\perp}^{(2)}(2\omega) = \chi_{surf,\perp\perp\perp} E_{pump,\perp}^2(\omega)$ ($\perp$ stays for the local normal to the surface), it is the straightforward result of the numerical time-domain based approach. The resulting second-harmonic pattern, calculated with this surface polarization approach appears on Fig. 3(b). While far field radiation signatures of both approaches possess certain similarities, the near field maps are quite different. Full hydrodynamic solution delivers much more information about the near field interaction.

Third harmonic

Bulk third-harmonic generation in metals has been considered by the introduction of nonlocal ponderomotive force, acting on electrons in gradient electromagnetic field [20]. Higher-harmonic generation at the boundaries (e.g. flat metal surfaces), based on the Sommerfeld free-electron models with subsequent solution of the Schrödinger equation in the Kramers-Henneberger accelerating frame were also developed [21] and are in good agreement with experimental data [22]. However, impact of complex geometries on higher-harmonic generation is much less studied, due to extra challenges in experimental verifications, and because of more complex theoretical treatment required.

The developed here method enables to address these challenges as well. The third harmonic field pattern, calculated by spectral frequency filtering, appears on Fig. 3(c). It possesses strong beaming characteristics in the forward scattering direction. In the case of spherical particles and phenomenologically introduced bulk third order nonlinearity nonlinear Mie theory predicts similar behaviours [23]. Fig. 3(d) shows the modelling results on third-harmonic generation, recovered from phenomenologically introduced bulk $\chi^{(3)}$. As in the case of SHG, the far-field patterns o these models agree with each other, while the near-field structure is better reproduced by full hydrodynamic model.

## 4. Outlook and Conclusions

Time-domain implementation of hydrodynamic model for conduction electron in metals, developed here, enables non-perturbative investigations of various nonlinear interactions between light and plasmonic nanostructures without additional approximations. While various typically used approaches make additional restricting assumptions on nonlinear dynamics (e.g. undepleted pump, uncoupled frequencies approximations) in order to simplify solutions of coupled nonlinear equations, the proposed method enabled to obtain universal, self-consistent numerical solution free from these approximations. Moreover, time-domain solutions enable the investigation of interactions with arbitrary shaped optical pulses (in contrast to generally employed CW pump assumption), opening the opportunity to approach realistic experimental scenarios. Arbitrary structures, resonant at certain frequencies (e.g. excitation, nonlinear scattering or both), can be comprehensively studied. The developed non-perturbative model enables investigations vast of multidisciplinary problems, involving metal composites interacting with weak, moderate or intense optical pulses.

**Acknowledgments**

This work has been supported, in part, by EPSRC (UK) and ARO (US). A.Z. acknowledges support from The Royal Society and the Wolfson Foundation. G.W. is grateful for support from the EC FP7 project 304179 (Marie Curie Actions).

**Figure Captions**

Fig. 1. (Colour Online) (a) Schematics of nonlinear optical generation from a gold cylinder, illuminated by laser beam (red). (b) Schematics of the simulation domain.

Fig. 2.(Colour Online) Nonlinear response of the system. (a) Nonlinear spectrum - Fourier transform of the time-dependent signal ($\alpha(\omega)$), at one point within the simulation domain. (b) Integrated radiated power of the second harmonic signal as the function of normalized and squared incident intensity. (c) Integrated radiated power of the third harmonic signal as the function of normalized and cubic incident intensity.

Fig. 3. (Colour Online) Nonlinear radiation patterns. Field intensity at the (a,b) second harmonic (c,d) third harmonic. Full hydrodynamic model (a,c), phenomenological approach (b,d)

**List of Figures:**

**Fig. 1**

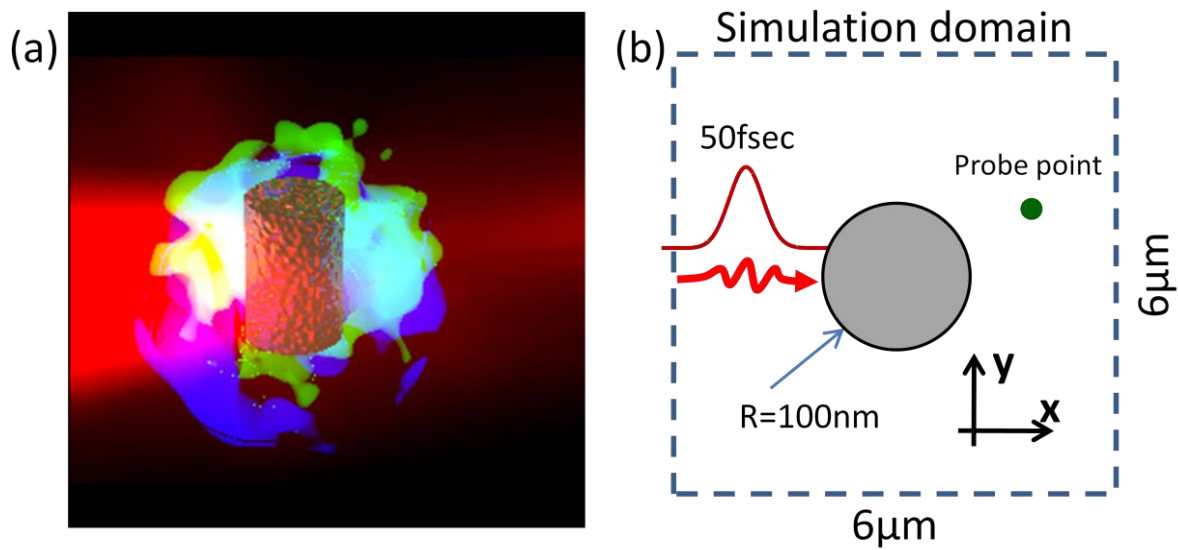

**Fig. 2**

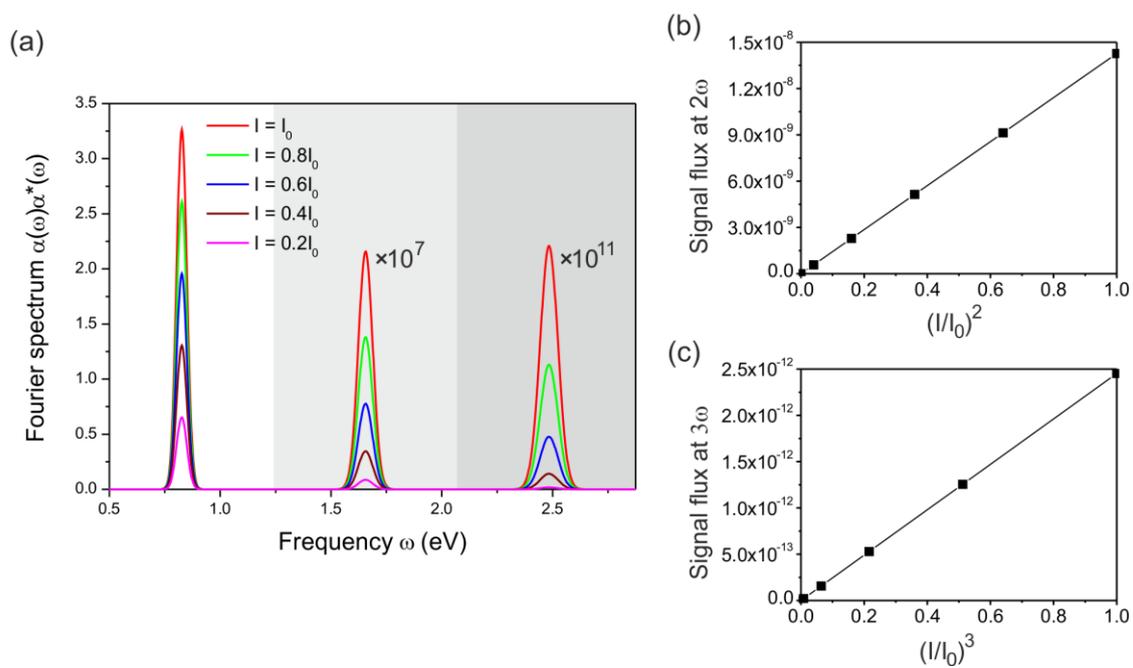

**Fig. 3**

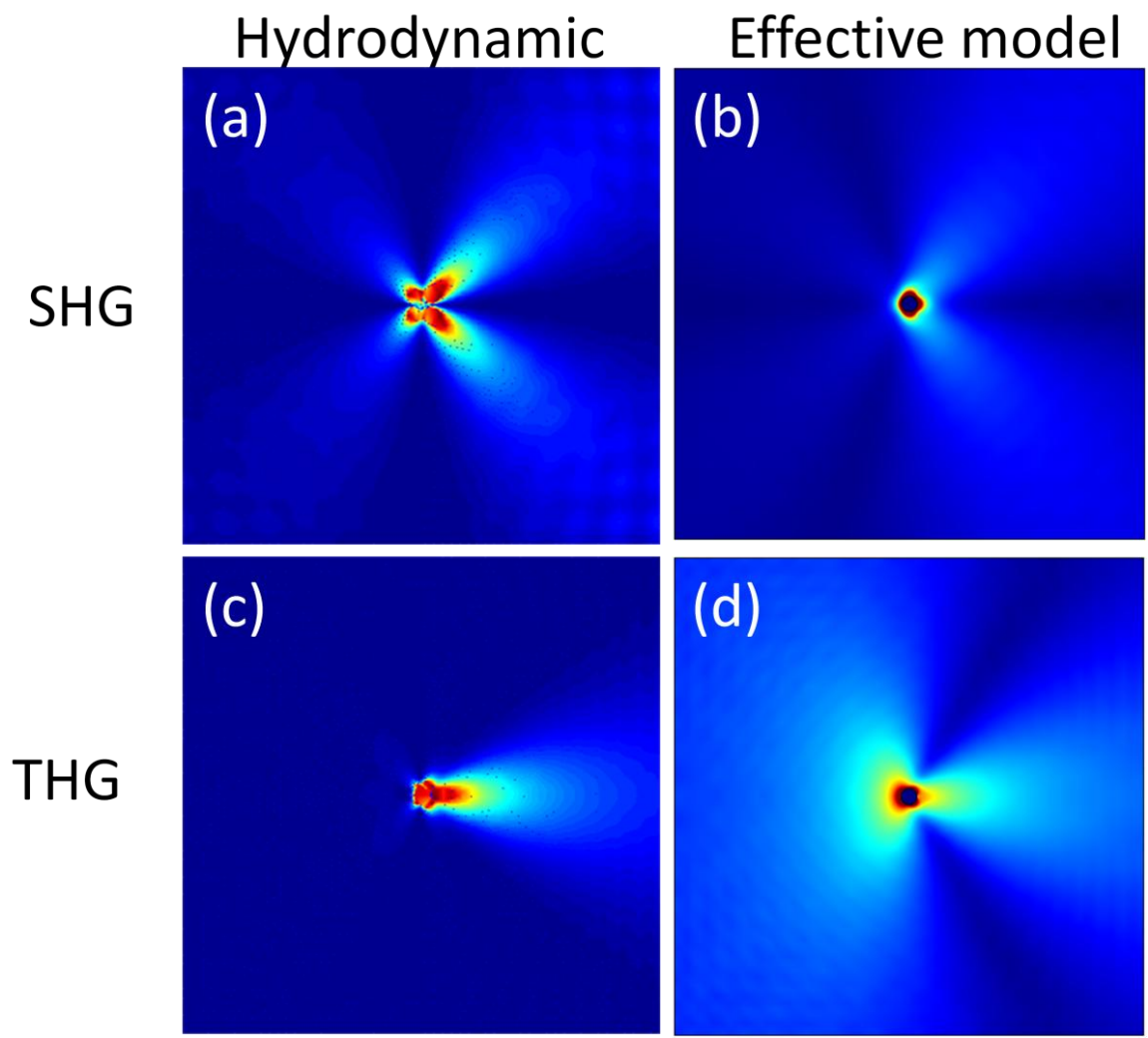